\documentclass[a4paper,prb,twocolumn,floatfix,showpacs]{revtex4-1}
\usepackage{graphicx}

\begin{document}


\title{Structure and magnetism of disordered carbon}

\author{M. A.~Akhukov}
\author{M. I.~Katsnelson}
\author{A.~Fasolino}

\affiliation{Radboud University Nijmegen/Institute for Molecules and Materials, Heyendaalseweg 135, NL-6525AJ Nijmegen, The Netherlands}
{a.fasolino@science.ru.nl}

\begin{abstract}
Motivated by the observation of ferromagnetism in carbon foams, a massive search for (meta)stable disorder structures of elemental carbon is performed by a generate and test approach.
We use the Density Functional based program SIESTA to optimize the structures and calculate the electronic spectra and spin densities.
About 1\% of the 24000 optimized structures presents magnetic moments, a necessary but not sufficient condition for intrinsic magnetic order.  We analyze the results using elements of graph theory. Although 
the relation between structure and the occurrence of magnetic moments is not yet fully clarified, we give some minimal requirements for this possibility, 
such as the existence of three-fold coordinated atoms surrounded by four-fold coordinated atoms. We discuss in detail the most promising structures. 

\end{abstract}


\maketitle

\section{Introduction}~

Carbon has many ordered allotropes but also interesting amorphous structures.
As a result of particular growth mechanisms it is found also in 
disordered porous structures formed by interconnected nanometer sized clusters, 
often called nanofoams. This type of structures have been imaged in 
Transmission Electron Microscope (TEM) experiments \cite{Rode1999, Rode2002}.
Carbon nanofoams are
obtained by high-repetition-rate laser ablation of 
a glassy carbon target in an ambient non-reactive Ar atmosphere.
Scanning Tunneling Microscopy (STM) reveals a mixed $sp^2$ $sp^3$ bonding
and curved graphite-like sheets. The Fourier Transform of STM images reveals 
clusters arranged with period of 5.6 \AA~ \cite{Rode1999, Rode2002, Rode2004}. 

A reason of interest for these structures is the presence of  
ferromagnetic behaviour up to 90 K reported in Ref. \cite{Rode2004} 
which raises questions about the mechanism for this phenomenon. 
These disordered structures might present magnetic moments related to 
undercoordinated atoms or to particular atomic arrangements. 
The purpose of this work is to investigate, by means of 
a large number of realizations of nanosized clusters, 
if some atomic arrangements appear recurrently and can give rise 
to magnetic moments and/or to magnetic order.Our analysis provides minimal criteria 
for the occurrence of magnetic moments and proposes several structures of defected carbon 
that might present ferromagnetic order. 

We first describe in section \ref{sec:generate} the procedure to generate and relax
in an automated fashion, series of disordered samples and 
the criteria to analyse this large set of results (about 24300 realizations)
according to total energy, coordination and magnetic moments. 
In our search we consider cluster structures periodically repeated to form a bulk. 
In this way we disregard the possibility of magnetism related to the surfaces. 

In section \ref{sec:search} we present a first screening of the magnetic properties 
that are the focus of our study. 

In section \ref{sec:energy} we examine 
the distribution of total energy, some structural properties and try 
to establish a kind of phase diagram to relate the presence of 
sizeable magnetic moments to the total energy of the structure. 

In section \ref{sec:structure} we set up a model, 
based on the graph theory to analyse the networks of bonds around a magnetic atom. 
This model has been used to analyse the allotropes of $C_{60}$ in a famous paper 
by Wales \cite{c60-funnel} revealing a distribution of energy with many deep minima separated 
by high barriers forming a funnel with minimum energy corresponding to the 
icosahedral $C_{60}$. Furthermore the graph theory can be related to the concept 
of bipartite lattices that play an important role in graphene.
Also in our search we find that carbon can form a wealth of 
metastable disordered structures with not a high penalty in terms of energy. 
We aim at finding the reason why some of these realizations carry magnetic moments 
and may even lead to ferromagnetic behaviour. 

In section \ref{sec:exchange-energy} 
we single out only the structures with sizeable magnetic moments and analyse them 
in the spirit of the mean field approximation in terms of exchange energies. 
Most magnetic structures have antiferromagnetic order but 
a few of them present a hint of ferri- or ferromagnetic order.
Unfortunately the LDA-CA and GGA-PBE approximations often do not agree in the evaluation 
of total energy and give small variations in bond lengths. Nevertheless we show
that besides these disagreements result obtained in both model are qualitatively consistent.

In the last section \ref{sec:examples} we focus on the most interesting samples 
found in our search and try to establish some recurrent features and 
minimal requirements for the presence of magnetic order. 

Our search and analysis is certainly not exhaustive and many questions 
remain open but it represents a first systematic attempt to grasp the physics 
and bonding leading to ferri/ferromagnetism in disordered carbon structures. 
We also 
give\cite{ac-xsf-data} the structure and coordinates of the 
selected samples presented in section \ref{sec:examples}
in the xsf file format suitable for use in the VESTA program \cite{vesta}.

\section{Procedure for the generation of disordered samples}~
\label{sec:generate}

In this section we describe how we generate nanosized disordered samples, 
relax them to find (meta)stable structures and 
then calculate their magnetic properties. 
    We calculate electronic and magnetic structures within the DFT \cite{DFT-1, DFT-2}
by means of the SIESTA code which implements DFT on a localized basis set 
\cite{SIESTA-1, SIESTA-2, SIESTA-3}.
    We used LDA with Ceperly and Alder parametrization (LDA-CA)
\cite{LDA-CA-1, LDA-CA-2} and a standard built-in double-$\zeta$ polarized (DZP)
\cite{DZP-NAO} basis set to perform geometry relaxation.
For some cases we used also the Generalized Gradient Approximation with Perdew-Burke-Ernzerhof exchange model (GGA-PBE) \cite{GGA-PBE}.
In general, the GGA gives much more accurate results for cohesive energy,
equilibrium structure and related characteristics of molecules and crystals.
Neither LDA nor GGA, however, can take into account
van der Waals interactions which are crucially important to describe
the interlayer binding in graphite.
As a result, GGA without van der Waals interaction cannot describe
the stability of graphite \cite{Rydberg2003-vdw-dft} whereas,
by chance, due to error cancellation, LDA gives a relatively accurate
interlayer distance and binding energy in graphite.
Therefore, it is now common practice to use the LDA for calculation of
multilayer graphitic systems (see e.g. \cite{Boukhvalov-bl-graphene} and references therein). 
The price to pay for this choice is that diamond becomes slighlty more energetically favourable than graphite (see Fig.\ref{fig:EtotPerAtom})
contrary to experiment. We do not think that this shortcoming is important for our results.
 
    The DZP basis set represents core electrons by norm-conserving
Troullier-Martins pseudopotentials \cite{pseudopotentials} in the
Kleynman-Bylander nonlocal form \cite{nonlocal-form}. For a carbon atom
this basis set has 13 atomic orbitals: a double-$\zeta$ for 2s and 2p
valence orbitals and a single-$\zeta$ set of five d orbitals.
    The cutoff radii of the atomic orbitals were obtained from an
energy shift equal to 0.02 Ry which gives a cut-off radius of 2.22 \AA~
for s orbitals and 2.58 \AA~ for p orbitals.
    The real-space grid is equivalent to a plane-wave cutoff energy of 400 Ry,
yielding $\approx$ 0.08 \AA~ resolution for the sampling of real space.
    We used k-point sampling of the Brillouin zone based on the Monkhorst-Pack
scheme \cite{Monkhorst-Pack} with 32 k-points.
    An iterative conjugate gradient (CG) procedure is then applied to reach 
stable or metastable structures.
The geometries were relaxed until all interatomic forces were smaller than 0.04 eV/\AA~ 
and the total stress less than 0.0005 eV/\AA$^3$.
    No geometrical constrains were applied during relaxation.
It is important to notice that, due to the random nature of the samples, many of them 
have to be metastable also after the CG minimization and could evolve after annealing by  molecular dynamics to energetically 
more favourable structure and change their magnetic property.

To construct structures similar to those observed experimentally 
\cite{Rode1999, Rode2002},
we generated samples with a given number of atoms from 5 to 64 in periodically repeated unit cells with size 5-10 \AA.
To simulate the experimental conditions of high pressure
we used geometries compressed up to 45\% of their equilibrium size. 
According to our calculations by DFT with the SIESTA code the initial structures 
have an internal pressure of 200-600 GPa.

To compensate the high internal stress, the CG relaxation
leads to drastic changes of the initial structure.
In this way, the minimization procedure gives a chance to reach  high energy 
metastable configurations with the possible presence of magnetic states, 
similar to the situation observed experimentally for carbon nanofoams.

\begin{table*}[htb]
    \caption{Number of magnetic samples with total (absolute) spin polarization 
             $m$ ($m_{abs}$) calculated using equations (\ref{eq:tsp})
             corresponding to a set of magnetization intervals in $\mu_B$.
             }
    \centering 
    \begin{tabular}{*{8}{c}} 
        \hline\hline 
        $N_{atoms}$ & $N_{samples}$ & $m(m_{abs})>0.01$ & $>0.05$ & $>0.10$ & $>0.25$ & $>0.5$ & $>1.0$ \\
        \hline 
         5  &  1000  &  22(23)  &  22(23)  &  21(22)  &   2(2)   &   2(2)   &  0(0)  \\
         6  &  1000  &   0(0)   &   0(0)   &   0(0)   &   0(0)   &   0(0)   &  0(0)  \\
         7  &  1000  &  26(26)  &  26(26)  &  26(26)  &  14(14)  &  14(14)  &  0(0)  \\
         8  & 10000  &   7(7)   &   3(7)   &   3(4)   &   1(2)   &   0(1)   &  0(1)  \\
         9  &  1000  &   7(9)   &   7(7)   &   6(6)   &   3(3)   &   3(3)   &  0(0)  \\
        10  &  4000  &  27(27)  &  27(27)  &  26(26)  &  22(23)  &  16(18)  &  0(0)  \\
        11  &  1000  &   6(6)   &   6(6)   &   6(6)   &   5(5)   &   3(3)   &  0(0)  \\
        12  &  1000  &  14(15)  &  13(14)  &  12(13)  &   9(11)  &   3(7)   &  0(2)  \\
        13  &  1000  &  11(13)  &  10(11)  &  10(11)  &   8(9)   &   6(7)   &  1(3)  \\
        14  &  1000  &  15(15)  &  13(14)  &  12(14)  &  11(13)  &   7(12)  &  0(2)  \\
        15  &  1000  &  16(16)  &  15(16)  &  14(16)  &  12(15)  &  10(15)  &  1(2)  \\
        16  &  1100  &  28(28)  &  24(28)  &  22(27)  &  18(25)  &  13(21)  &  0(5)  \\
        24  &   100  &   4(4)   &   4(4)   &   4(4)   &   4(4)   &   3(4)   &  0(0)  \\
        64  &   100  &   5(13)  &   3(9)   &   3(8)   &   3(8)   &   3(8)   &  3(8)  \\
        \hline 
    \end{tabular}
    \label{tab:mag} 
\end{table*}

To generate the initial geometries we used the following approach.
First of all, we calculate the volume per atom in the graphite unit cell:

\begin{equation}
\label{volume-per-atom-in-graphite}
v^{graphite}_{atom} = \frac{3 \sqrt{3}} {4} r_{cc}^2 r_{ll}
\end{equation}

where $r_{cc} = 1.42$ \AA~ is the carbon-carbon interatomic distance in the layer
and $r_{ll} = 3.35$ \AA~ is the interlayer distance in graphite.
Since we will construct compressed unit cells by scaling of the coordinates, 
it is convenient to express $r_{ll}$ in terms of $r_{cc}$. 
For graphite  $r_{ll} = 2.36~ r_{cc}$.
In this way, the volume $v^{graphite}_{atom}$ can be written as proportional to $r_{cc}^3$
and 
$\sqrt[3]{v^{graphite}_{atom}}$ becomes  proportional to $r_{cc}$.
Rescaling the unit cell size by $r_{cc}^{custom}$ we can construct the
initial cubic unit cell for a given number of atoms $N_{atoms}=n$
with lattice constant $a$ as

\begin{equation}
a =  \frac{\sqrt[3]{n v^{graphite}_{atom}}} {r_{cc}} r_{cc}^{custom}
\end{equation}
    
To allocate the required number of atoms within the prepared cubic unit cell 
we randomly generated atomic coordinates. 
If the newly generated position is closer than 
$r_{cc}^{custom}$ to any atom we replaced such a pair 
by one atom with average position. 
At the end, the geometry has no atoms closer than $r_{cc}^{custom}$. 
For the case  $r_{cc}^{custom}$ = 1.42 \AA~ the density 
of the generated system is equal to the one of graphite. 
We used $r_{cc}^{custom}$ = 1.1 \AA~ that gives a density 
$\approx$2.15 higher than graphite.


\begin{table*}[ht]
    \caption{
        An overview of the obtained information about 
        coordination in all studied configurations.
        The percentage of samples with specific coordination 
        is shown for each series of atoms.
        Since the case of 6 3-fold atoms is not applicable to the series with 5 atoms 
        and it duplicates the value of all 3-folds configurations for the 
        series with 6 atoms we keep these two cells empty.}
    \centering 
    \begin{tabular}{c c c c c c c c c c} 
        \hline\hline 
        $N_{atoms}$ & $N_{samples}$ & all 3-fold & 2 3-fold & 4 3-fold & 6 3-fold & all 4-fold & 1 2-fold & 2 2-fold & 3 2-fold  \\
        \hline 
           5  &  1000  &   0.0  &  15.4  &  38.1  &   -   &  46.4  &  17.8  &  0.0  &  1.3 \\
           6  &  1000  &  33.2  &  32.4  &  25.0  &   -   &   9.3  &   7.3  &  4.9  &  0.0 \\
           7  &  1000  &   0.0  &  30.8  &  19.1  & 32.2  &  17.3  &   8.1  &  0.2  &  0.0 \\
           8  & 10000  &  22.7  &  22.0  &  23.6  &  9.0  &  22.4  &   1.5  &  0.4  &  0.0 \\
           9  &  1000  &   0.0  &  25.6  &  31.3  & 14.3  &  17.2  &   1.8  &  0.1  &  0.0 \\
          10  &  4000  &  10.2  &  22.5  &  22.5  & 24.1  &  11.3  &   2.5  &  0.6  &  0.05\\
          11  &  1000  &   0.0  &  23.0  &  29.6  & 19.9  &  12.8  &   3.2  &  0.5  &  0.2 \\
          12  &  1000  &   4.6  &  21.1  &  24.7  & 16.7  &  12.8  &   1.8  &  0.3  &  0.0 \\
          13  &  1000  &   0.0  &  24.1  &  24.1  & 19.5  &   7.1  &   3.0  &  0.5  &  0.0 \\
          14  &  1000  &   2.0  &  17.3  &  21.8  & 23.9  &   6.5  &   2.6  &  0.4  &  0.0 \\
          15  &  1000  &   0.0  &  18.0  &  21.6  & 22.8  &   6.5  &   4.5  &  1.3  &  0.0 \\
          16  &  1100  &   1.2  &  16.3  &  21.5  & 20.7  &   7.5  &   4.5  &  0.3  &  0.2 \\
          24  &   100  &   0.0  &   4.0  &  11.0  & 21.0  &   0.0  &   6.0  &  3.0  &  0.0 \\
          64  &   100  &   0.0  &   0.0  &   0.0  &  0.0  &   0.0  &  11.0  &  4.0  &  1.0 \\
        \hline 
    \end{tabular}
    \label{tab:coordination} 
\end{table*}

Such a randomly generated geometry with high internal pressure and fixed
number of atoms per unit cell is then minimized by CG 
letting the atomic positions within the cell and the cell lattice parameters vary. 
We then study each sample to search for magnetic states in pure carbon materials. 
We have studied in this way 24300 samples. 
A first screening gives $\approx$1\% (see Table \ref{tab:mag}) of the samples 
with magnetic states.
In Table \ref{tab:coordination} we report the coordination in samples 
with 5 to 64 atoms in the unit cell, calculated automatically by counting the number of neighbors closer than 1.8~\AA.  

In the following we use the following notation to refer to a specific sample, 
namely ACxx-yyyy where AC stays for amorphous carbon, 
xx is the number of atoms in the unit cell and yyyy denotes a specific sample. 
As an example, the sample AC07-0010 shown in Fig. \ref{fig:AC07-0010} 
is the tenth of a series with 7 atoms in the unit cell.

We see that also samples with only 3-fold or 4-fold bonding are found.  
Their number decreases for the larger unit cells as expected because 
there are many more possible configurations. 
We find, as expected from considerations that will be discussed in section \ref{sec:structure}, 
no samples with all 3-fold atoms for odd $N_{atoms}$. 
Unexpected is that for $N_{atoms}=8$ there is a maximum number of all 4-fold samples.
For the samples with mixed bonding we report the percentage of samples 
with 2, 4 or 6 3-fold bonded atoms. In general the 4-fold atoms have bonding angles very close to that of $sp^3$ 
hybridization in diamond and very often 3-fold atoms are almost flat $sp^2$ configurations like in  graphitic forms of carbon.
Within the many possible configurations, we distinguish those with 
all 3-fold atoms (graphite-like $sp^2$ structure), all 4-fold atoms (diamond-like $sp^3$ structure),
configurations with 2-fold coordinated atoms and
mixed configurations with different percentage of 3-fold and 4-fold atoms.
The two first groups (all 3-fold and all 4-fold) are non magnetic and the last two groups
are possible candidates for magnetic states as it will be discussed in more details in
section \ref{sec:structure}. 

To distinguish between ferromagnetic and antiferromagnetic samples we use the
total spin polarization $m$  and the total absolute spin polarization $m_{abs}$ 

\begin{equation}
\label{eq:tsp}
m = \sum_{i=1}^{N_{Atoms}} s_i,
~~~~
m_{abs} = \sum_{i=1}^{N_{Atoms}} |s_i|,
\end{equation}

\begin{equation}
s_i = q_i^{up}-q_i^{down}
\end{equation}

where $q_i^{up}$  is the charge corresponding to spin ''up'' at the i-th atoms and
$q_i^{down}$ is the charge corresponding to spin ''down'' at the i-th atoms.

We can then distinguish  antiferromagnetic samples, 
where  $m = 0$ and $m_{abs} \neq 0$,
and ferromagnetic ones for which $m = m_{abs}$.
Moreover we found a set of samples where $m > 0$ and $m \neq m_{abs}$.
This is a sign of ferrimagnetic properties.

\section{Search of magnetic states}~
\label{sec:search}

To identify the geometrical structures responsible for the magnetic states, 
we perform a numerical experiment based on a {\it generate and test}  approach 
\cite{generate-test}
with elements of genetic algorithms 
\cite{genetic-algorithm}. 
Such a method is convenient in view of the available large 
amount of computational facilities 
which allows to calculate automatically thousands of independent configurations. 
We varied the number of atoms together with the  unit cell size and number of 
configurations in the computational series iteratively 
to identify the most typical geometrical structures carrying magnetic states. 

We started with 64 atoms per sample and 100 samples in the series. 
We found 5 configurations with total spin polarization 
$m>0.010 \mu_B$ and only 3 with $m>0.500 \mu_B$,
(see Table \ref{tab:mag} last row). At the same time, from Table \ref{tab:mag},
we see that within the series with 64 atoms besides 
the samples with sizeable total spin polarization $m$
there are a number of samples (i.e. 8-3=5 samples) where $m$ 
is small while $m_{abs}$ is of the order of $\mu_B$.
This facts is a clear evidence of antiferromagnetic arrangement of magnetic moments 
in the ground state as discussed at the end of section \ref{sec:generate}.

\begin{figure}[htb]
    \centering
    \includegraphics[width=0.50\textwidth]{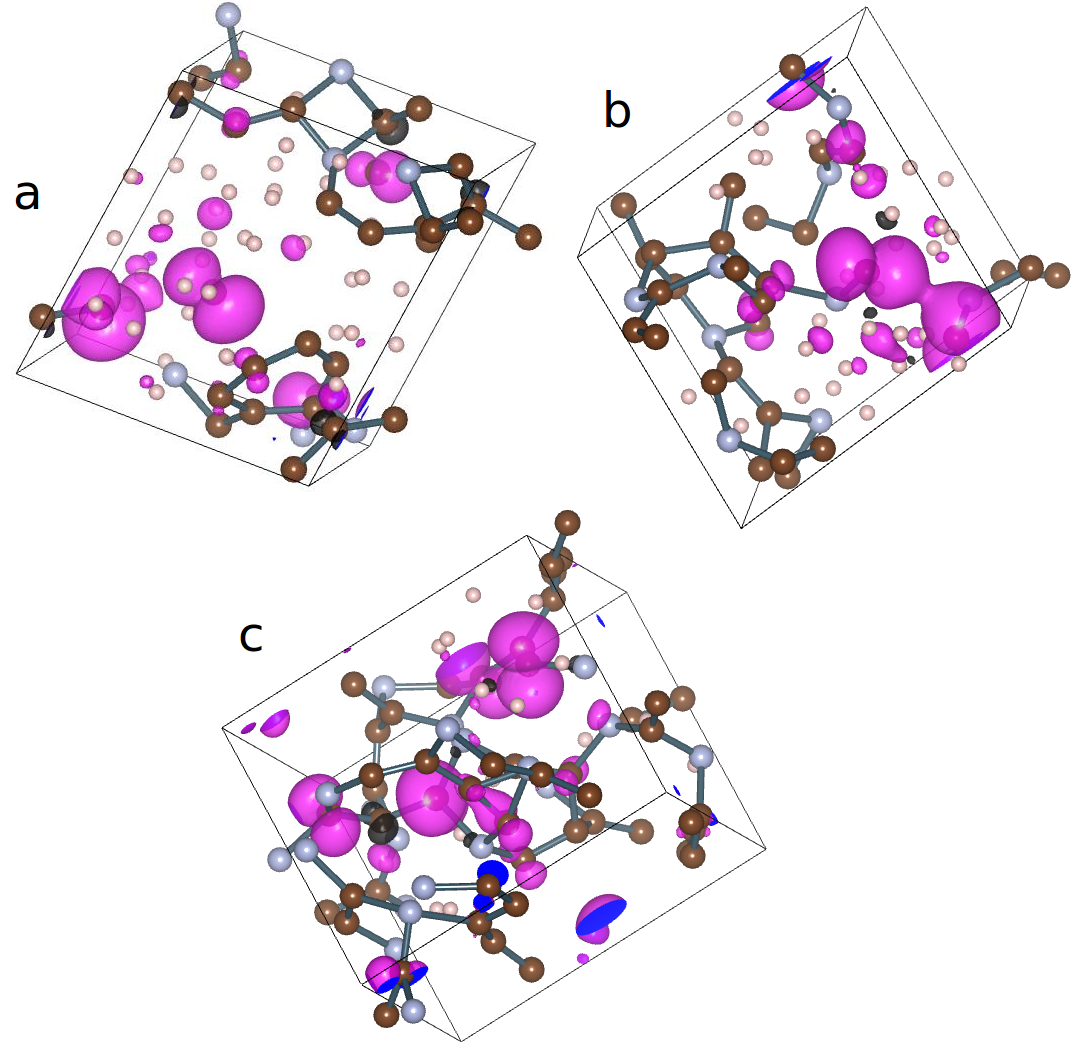}
    \caption{
        Examples of spin polarized structures for
        disordered carbon (64 atoms per unit cell) after relaxation, 
        showing atoms with magnetic states (pink clouds) 
        surrounded by nonmagnetic ones. 
        The 4-fold atoms are small balls marked by light gray, 
        the 3-fold atoms are balls marked by brown. 
        The 4-fold with 3-fold neighbours are balls marked by gray. 
        The 1D chains of 3-fold atoms separated by single 4-fold atoms are marked by bonds. 
        Here we see well distinguishable networks made of 1D chains of 3-fold atoms.
        We see that not all the magnetic atoms belong to these chains.
        The geometry is visualized by the VESTA program \cite{vesta}.
    }
    \label{fig:AC64}
\end{figure}

A first analysis shows that magnetic moments are carried by 
individual atoms in specific atomic configurations. 
Typical configurations with magnetic atoms are shown in 
Fig. \ref{fig:AC64}a,b,c. We have highlighted  the presence of networks of 3-fold atoms. 
We will come back to the relation between these networks and 
the magnetic atoms in section \ref{sec:structure}.

One might have expected the source of uncompensated spin to be 
dangling bonds originating from 2-fold carbon atoms as it was shown for grain boundaries in 
\cite{DBGB}. We find instead that also 3-fold carbon atoms may have 
uncompensated spin and that this latter case occurs at least one order of 
magnitude more frequently than 2-fold atoms. 
While for 2-fold coordinated atoms the magnetic moment is clearly related to a dangling bond, 
the situation of 3-fold coordinated atoms is more complex. 
Depending on the bond angles, it may correspond to a planar $sp^2$ configuration 
like in graphite or to a $sp^3$ configuration with a dangling bond 
as found for instance at the ideal (111) diamond surfaces. 

Within the first series of samples with 64 atoms 
we have shown that only 3 configurations have more than one magnetic atom. 
This suggests that probably the geometrical conditions which make atoms
magnetic  can be detected 
in smaller and simpler geometries with fewer atoms per unit cell. 
To do so we iteratively reduced the number of atoms together with the unit cell size.

In the second series with 100 configuration and 
24 atoms per sample we found 4 configurations with 
total spin polarization $m>0.005 \mu_B$ and only 3 with $m>0.500 \mu_B$. 
Typically, we had 1-2 magnetic atoms per unit cell. 

In view of the small number of magnetic atoms in each sample, 
to make our search for structures with magnetic states more efficient we  generated and optimized series of thousands of 
configurations with less atoms (16 to 5 atoms) per unit cell. 
Let us  notice a few interesting facts.  First of all, we did not find 
any magnetic configurations within 1000 samples in the series with 6 atoms 
per unit cell and only 3 magnetic configurations with $S>0.100 \mu_B$ 
within 10000 samples in the series with 8 atoms per unit cell. 
The series with  5 and 7 atoms per unit cell instead presents magnetic configurations. 
This could be a sign of the importance of the parity of the number of atoms in the unit cell. 

By reducing the  number of atoms per unit cell we also reduce 
the number of possible relaxed configurations. This reduction 
is reflected in the presence of duplicate geometries up to 
variations in the directions of the unit cell vectors. 
The presence of duplicate geometries also suggests that, within the constraints
imposed by the sample construction, 
such geometries are more preferable than the others.

\section{Energy and magnetism}~
\label{sec:energy}

%

\begin{table*}[htb]
    \caption{
        Comparison of the 6 computational models used to calculate 
        the formation energy of 5 carbon allotropes, 
        i.e. 3 sets of parameters within GGA-PBE 
        and  3 sets of parameters within the LDA-CA approximation.
        Here $E_{mesh}$ gives the  resolution for the sampling of real space
        based on the plane-wave cutoff energy. 
        $E_{cutoff}$ is the cutoff radius of the atomic orbitals.
        (see section \ref{sec:generate})
        Two types of basis sets were chosen: the standard DZP and 
        the custom one constructed for graphitic materials using the approach
        described in \cite{DZP-NAO, variational-approach}. 
        Notice that the minimal energy structure is graphite
        for GGA-PBE and diamond for LDA-CA.}
    \centering 
    \begin{tabular}{c c c c} 
        \hline\hline 
          GGA-PBE, 32 k-points & DZP                  & DZP                   & custom \\
          $E_{mesh}$ = 400 Ry  & $E_{cutoff}$ = 1 mRy & $E_{cutoff}$ = 20 mRy & basis  \\
        \hline 
             diamond           &   0.021   &   0.008   &   0.112   \\
             graphene          &   0.007   &   0.066   &   0.003   \\
             graphite-A        &   0.002   &   0.015   &   0.001   \\
             graphite-AB       &   0.000   &   0.001   &   0.000   \\
             graphite-ABC      &   0.001   &   0.000   &   0.001   \\
    
        \hline\hline 
          LDA-CA, 32 k-points  & DZP                  & DZP                   & graphite \\
          $E_{mesh}$ = 400 Ry  & $E_{cutoff}$ = 1 mRy & $E_{cutoff}$ = 20 mRy & basis    \\
        \hline 
             diamond           &   0.000   &   0.000   &   0.000   \\
             graphene          &   0.195   &   0.257   &   0.079   \\
             graphite-A        &   0.173   &   0.166   &   0.059   \\
             graphite-AB       &   0.156   &   0.131   &   0.044   \\
             graphite-ABC      &   0.157   &   0.132   &   0.044   \\
        \hline 
    \end{tabular}
    \label{tab:models} 
\end{table*}

\begin{figure}[htb]
    \centering
    \includegraphics[width=0.50\textwidth]{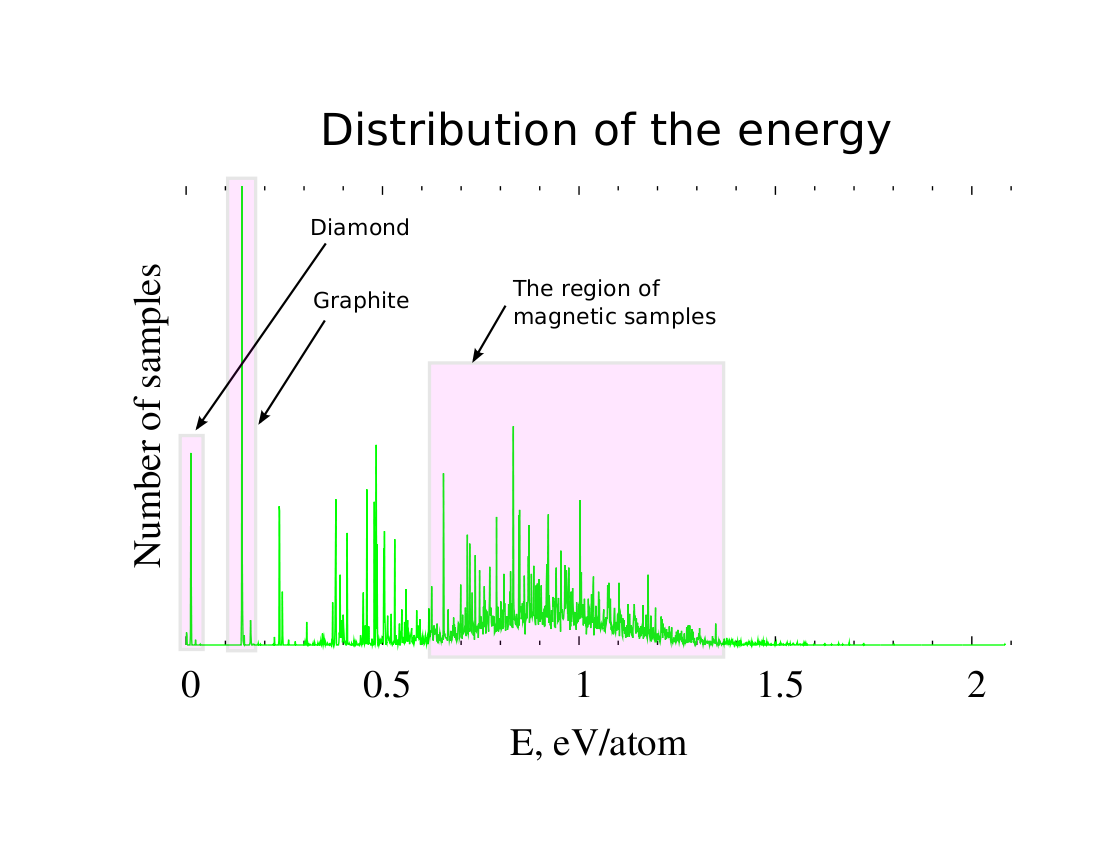}
    \caption{
        Distribution for the energy referred to diamond
        of all the calculated samples.
        The energy intervals marked as "diamond" and "graphite" 
        corresponds to the specified systems, 
        the third shaded energy interval contains
        magnetic samples and is further analysed in  Fig. \ref{fig:EtotStot}.
        }
    \label{fig:EtotPerAtom}
\end{figure}

\begin{figure*}[htb]
    \centering
    \includegraphics[width=0.9\textwidth]{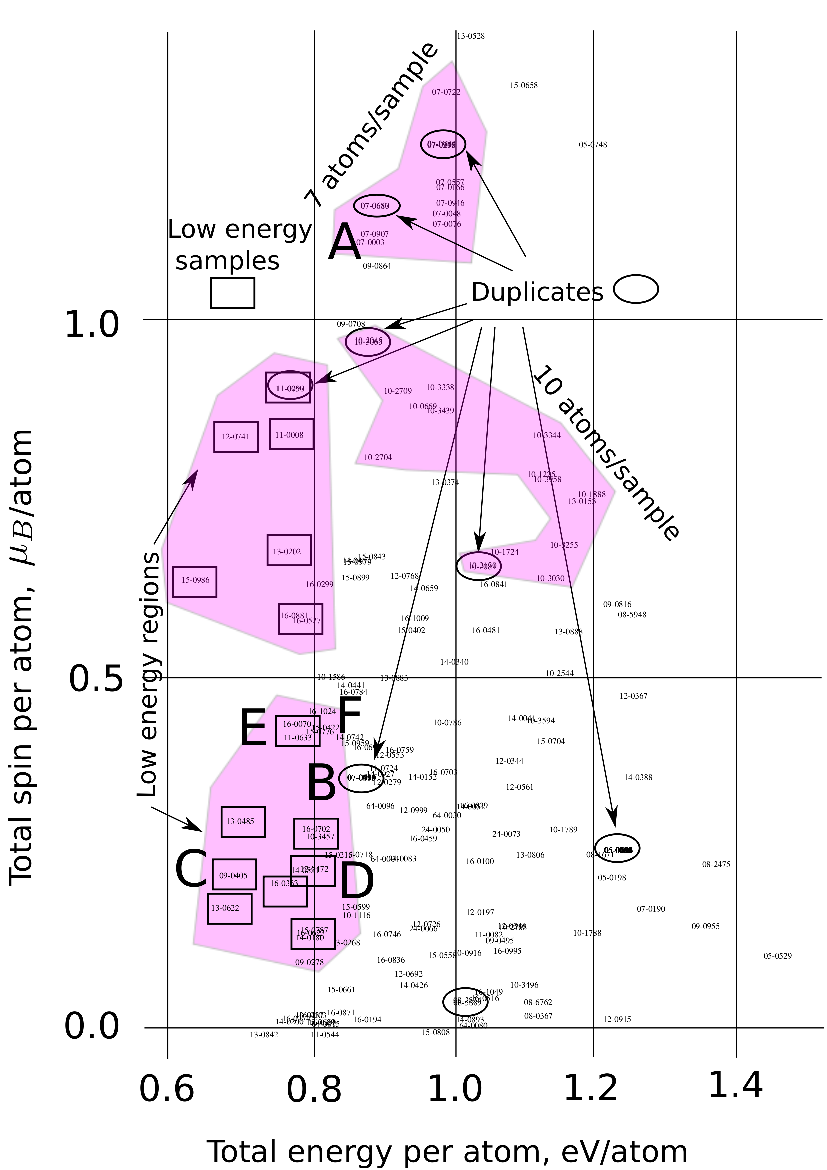}
    \caption{
        Total spin per atom as a function of total energy 
        per atom (referred to diamond) for magnetic structures. 
        Each point has a name combining the number of atoms and index in the series. 
        Few regions of spin and energy are marked by light grey polygons. 
        Within them, ellipses indicate the existence of many duplicates as described in the text. 
        The two left shaded regions contain low energy structures marked by rectangles.
        The structures indicated as A, B, C, D are shown 
        in Fig. \ref{fig:AC07-0003}, \ref{fig:AC07-0010},
        \ref{fig:ac-p2} right and \ref{fig:ac-p2} left respectively.
        }
    \label{fig:EtotStot}
\end{figure*}

Within thousands of calculated configurations with formation energy
lying between 0.0 and 2.0 eV/atom above diamond, as shown in Fig. \ref{fig:EtotPerAtom}
a few hundreds magnetic ones were found with different energies and spin polarizations. 
We note that, besides diamond and graphite there is another sharp peak at low energy.
Since these configurations are not magnetic we have not analysed them in detail but 
in a few cases we have seen that they relax to diamond or graphite after annealing. 
In Fig \ref{fig:EtotStot} we analyse the relation between formation energy and magnetic moments.
The total energy per atom for magnetic configurations lies in 
a relatively restricted range of 0.6-1.5 eV/atom above diamond and graphite,
most of them being in 0.8-1.2 eV/atom energy range.
In general we cannot identify any specific energy-spin region 
for particular series of calculations 
excluding the series with 7 atoms per sample (see top polygon in Fig. \ref{fig:EtotStot}) 
and partly the series with 10 atoms per sample (see middle polygon in Fig. \ref{fig:EtotStot}).
The presence of duplicates, especially for the series with 5-11 atoms 
that we have already discussed in \ref{sec:search}, is indicated by ellipses.
Low energy configurations are indicated by rectangles and we see that they cluster
in the two regions indicated by shaded polygons for energy $\le$ 0.8 eV/atom, 
only 10-15 \% higher than diamond and graphite.

The special character of the energy landscape of carbon systems, 
related to the fact that carbon can form a wealth of structures, 
often with very high energy barriers among them, has been pointed out 
in the seminal work \cite{c60-funnel} where all the 1812 
possible structures of $C_{60}$ are shown to form a funnel of well separated, 
deep minima not too far in energy from the absolute minimum 
corresponding to icosahedral $C_{60}$.

\section{Structure and magnetism}~
\label{sec:structure}

As we have seen in the previous section, magnetic structures are rare and 
the relation between structure and magnetism is certainly not trivial. 
We try and use the graph theory \cite{graph-theory} to relate structure and magnetism.
To take advantage from this theory we need to have only one odd numbered coordination. 
This is realized in our carbon system where no atoms with one bond  or five bonds are present 
because such configurations are especially unstable and were never observed in our samples 
after geometrical relaxation.

\begin{figure*}
    \centering
    \includegraphics[width=0.70\textwidth]{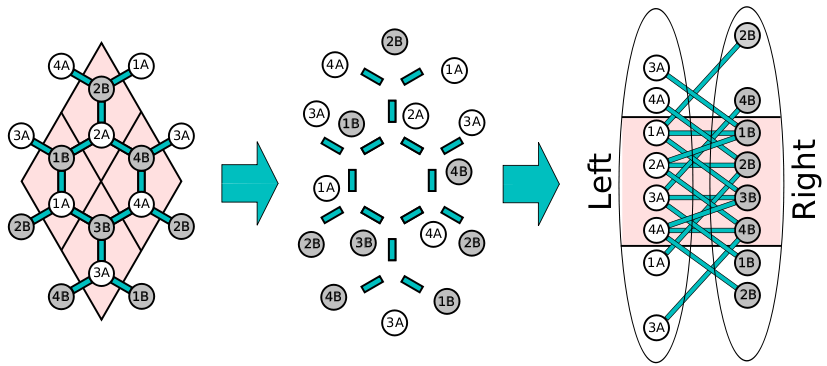}
    \caption{
        An example of the division procedure that maps the bipartite lattice of 
        graphene into two disjointed subgraphs called as Left and Right. 
        From left to right: a 2X2 supercell of graphene 
        indicated by the light grey polygons. 
        Atoms belonging to the A-sublattice (B-sublattice) are represented by 
        white (grey) balls with labels 1A(B), 2A(B), 3A(B), 4A(B).
        Since we consider 2D periodical structure we also show the 
        periodical images outside the unit cell. 
        Through an intermediate step we can construct the 
        Left and Right subgraphs as indicated by bringing all 
        A atoms to the left and all B atoms to the right 
        while keeping the bonds between former nearest neighbours. 
	All the atoms originally found in the shaded (pink) unit cell 
	are located in the shaded (pink) region of the graph.}
    \label{fig:bp-graph}
\end{figure*}

We start by a lemma that we have derived ourselves 
although it might be reported already in the literature.

{\bf Lemma}. Any connected graph with nodes having 2, 3 or 4 neighbours 
 always has an even number of nodes with 3 neighbours.

{\bf Proof}. Suppose we have $N_2$ nodes with 2 neighbours, 
$N_3$ nodes with 3 neighbours and $N_4$ nodes with 4 neighbours. 
Each edge connects two nodes. Due to this fact, the total number of edges 
in the graph is $(2 N_2 + 3 N_3 + 4 N_4)/2 = M$, 
where $M$ is an integer positive number, whence
$3 N_3 = (2 M - 4 N_4 - 2 N_2) = 2 (M - 2 N_4 - N_2)$.
This means that $3 N_3$ is an even number which implies that $N_3$ is even.

\begin{figure}
    \centering
    \includegraphics[width=0.50\textwidth]{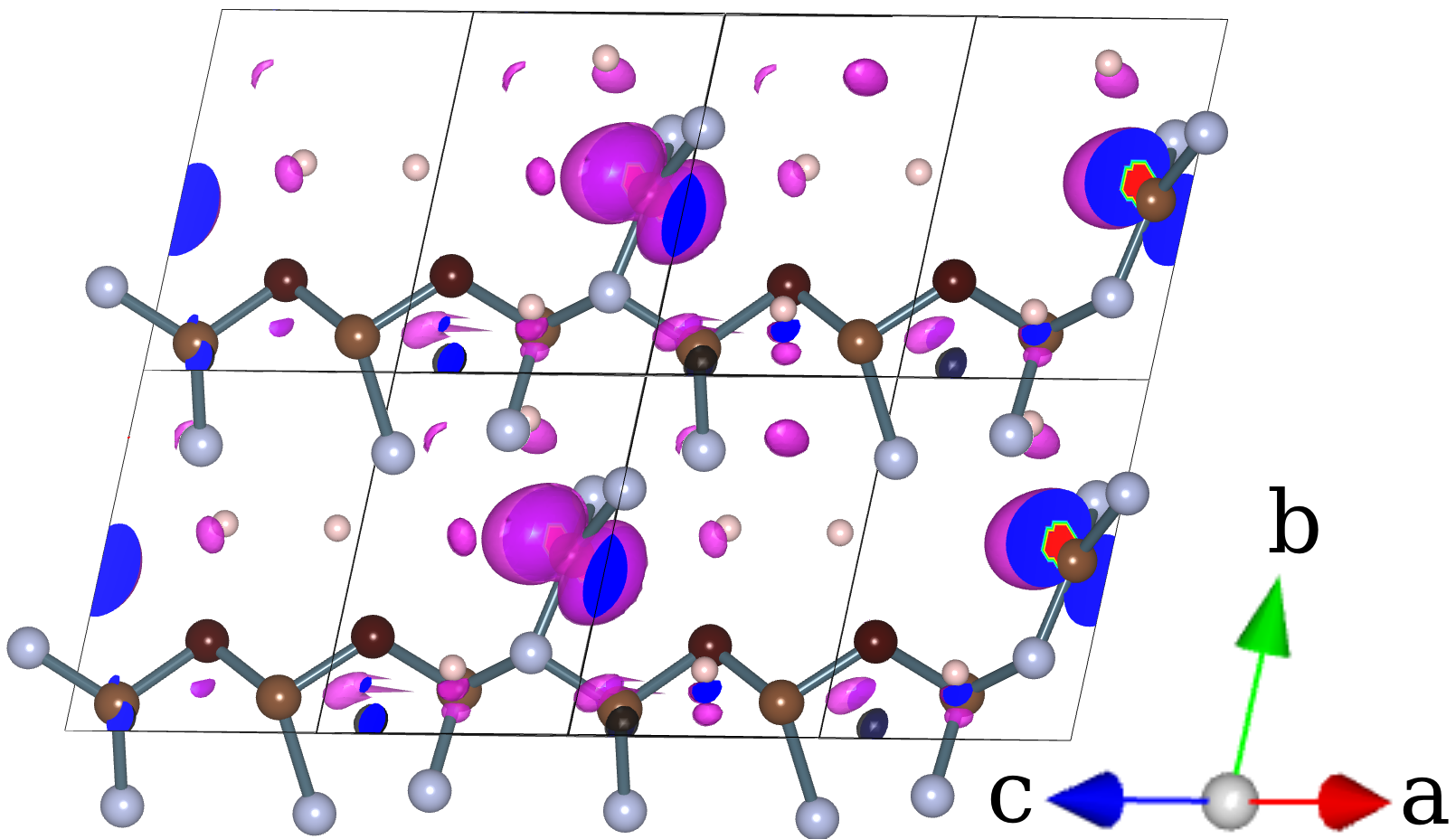}
    \includegraphics[width=0.05\textwidth]{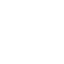}
    \includegraphics[width=0.50\textwidth]{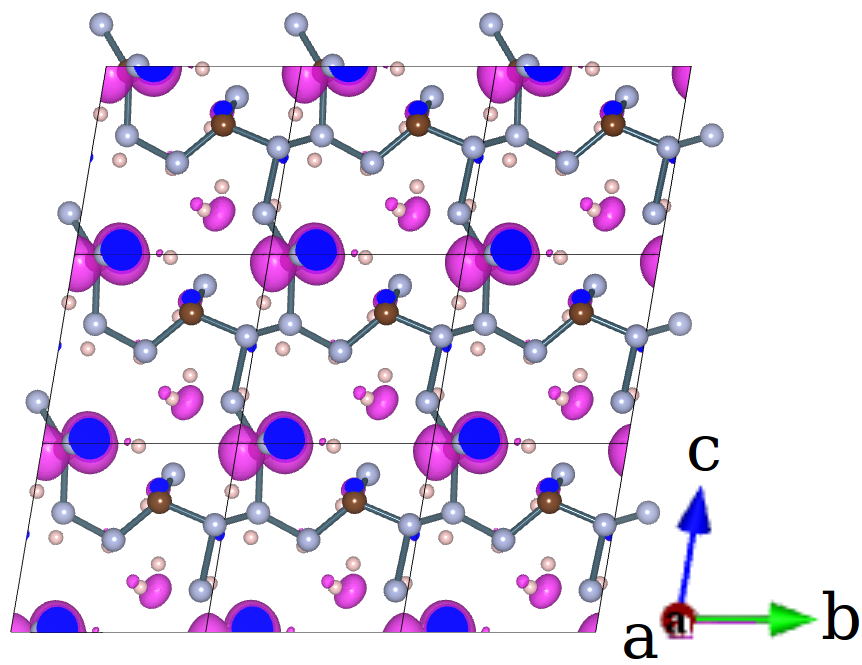}
    \caption{
	Examples of 1D chains (top) and 2D-networks (bottom) of connected 3-fold and 4-fold atoms.
        As previously (see Fig. \ref{fig:AC64}) we use different colours for 3-fold and 4-fold atoms with
        different neighbouring atoms.
        The 4-fold atoms are small balls marked by light grey, 
        the 3-fold atoms are balls marked by brown. 
        The 4-fold having 3-fold neighbours are balls marked by grey. 
}
    \label{fig:1D-chain-2D-web}
\end{figure}

The structure of the graph described in the lemma corresponds 
to the structure of disordered carbon 
where we have atoms with either 2, 3 or 4 neighbours. 
This case corresponds either to atoms in $sp$, $sp^2$ and $sp^3$ hybridization 
or atoms in $sp^2$ and $sp^3$ hybridization with 
dangling bonds (2 and 3 neighbours respectively).

The importance of this lemma lies, as it will be shown later, in the relation 
between magnetic states and the presence of atoms in $sp^2$ hybridization.
In principle, the simplest carrier of magnetic moments are dangling bonds on 2-fold coordinated atoms. This situation however, is rather fragile because these dangling bonds 
give a high energy penalty and will tend to be passivated in realistic situations. For this reason we need to focus on magnetic moments associated to 3-fold coordinated atoms, 
which represent a substantial fraction of the cases we have found.
We often observe a non compensated spin located 
at atoms in $sp^2$-hybridization with 3 neighbours and an unpaired $\pi$-orbital.
Another important observation is that $sp^2$ atoms usually group in 
network structures made of 1D chains of $sp^2$ atoms as shown in 
Fig. \ref{fig:AC64}.
Such a structure may be represented as a bipartite graph.
Graphene has a 2D bipartite unit cell \cite{Novoselov2004}.
An example of a graph 
representation for graphene is shown in Fig. \ref{fig:bp-graph}. 
According to the Lieb theorem \cite{Lieb-theorem, Yazyev2010-report, KatsnelsonBook} 
we can expect the presence of nonzero spin if the number of atoms 
in the Left and Right subgraphs is different. 
According to the above lemma, we always have an even number of atoms 
with 3 neighbours. It turns out that in most of the studied  structures this even number of 3-fold atoms is equally distributed in the Left and Right subgraphs. 
According to Lieb theorem in this case the ground state should be a non magnetic singlet state. 
As a result, to have magnetic states, we need specific geometric structures. 
For example, in the simplest case, the source of uncompensated spin 
could be a single $sp^2$ atom surrounded by $sp^3$ atoms. 
But, according to the lemma, we will always have at least one more $sp^2$ atom 
located somewhere in the unit cell.
If these two atoms are either in different subgraphs or form a bond, 
no magnetic states may be expected.

One straightforward consequence of the above considerations is 
that the absence of 3-fold atoms in the unit cell, 
i.e. a fully 4-fold structure, means the absence of any magnetic states. 
Within 24300 optimized configurations we have 4132 fully 4-fold structures 
and all of them are non magnetic. 

Since 3-fold atoms are the crucial ingredient to have magnetic states, 
we analyse our samples by evidencing the connections between 3-fold atoms. 
In this way we identify networks of bonds between them, 
as shown in Fig \ref{fig:AC64}a,b,c and more in detail in Fig \ref{fig:1D-chain-2D-web}.
To make the network of 3-fold atoms more evident 
we have marked the 4-fold atoms with tiny balls so that 
the remaining geometry looks like a 3D network of 1D chains of 3-fold atoms.
Such a 3D network may consist of a set of isolated clusters or may form an infinite structure 
due to the periodic boundary conditions in any or all directions of the unit cell.

\begin{figure}[htb]
    \centering
    \includegraphics[width=0.50\textwidth]{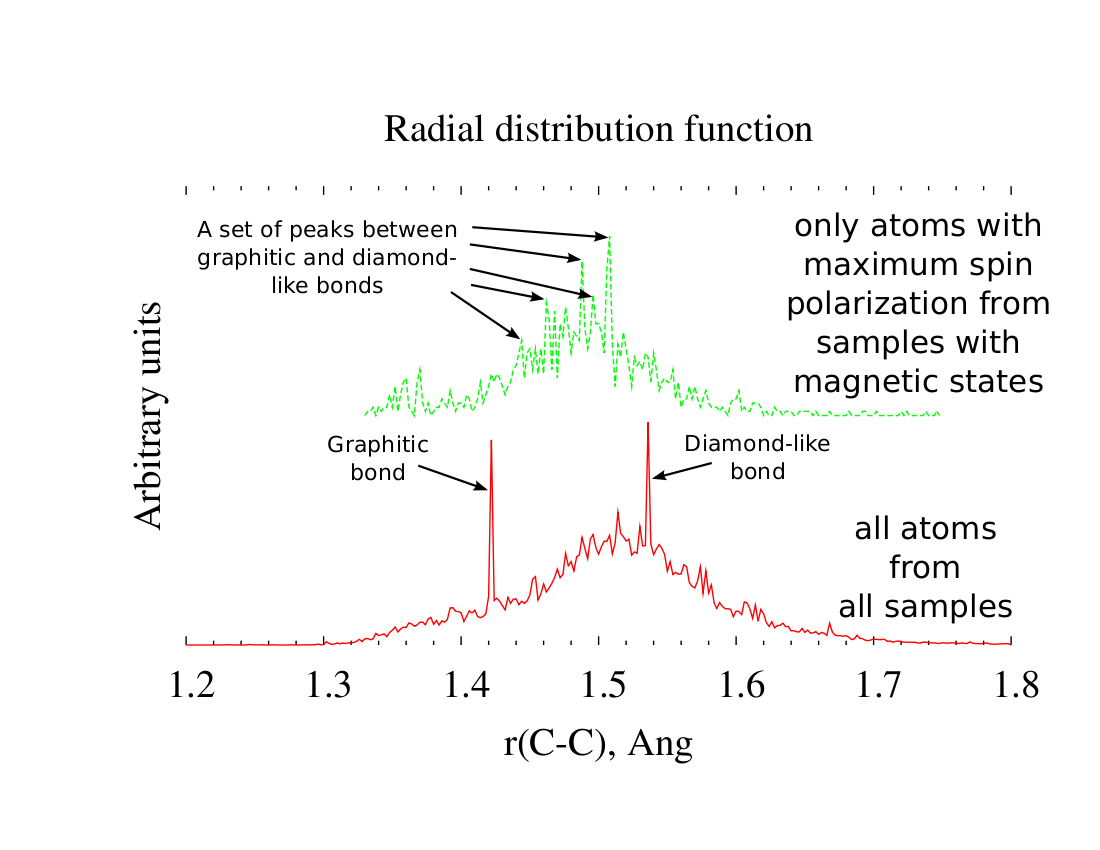}
    \caption{
        Comparison of the radial distribution functions 
        calculated for all atoms of all 24300 optimized samples 
        with that calculated by considering, for each sample with $m \neq 0$,
	only the atom with maximal magnetic moment.}
    \label{fig:rdf}
\end{figure}


Let us now consider one of the simplest cases where 
we can expect magnetic states on the basis of the lemma:  
two 3-fold atoms not bonded to each other within the unit cell.
We can automatically detect all such configurations. 
Within 24300 we found 990 samples with only two 3-fold atoms not bonded to 
each other where 34 of them have appreciable magnetic moments in 
ferro- or ferri- or antiferromagnetic configurations 
($m >$ 0.020 $\mu_B$ or $m_{abs} >$ 0.020 $\mu_B$). 
For completeness, we give the distribution of magnetic samples within each series:
8:1, 9:1, 10:6, 12:6, 13:3, 14:5, 15:4, 16:8 ($N_{atoms}$:$N_{samples}$)
where 3 of them are with $m < $0.020 $\mu_B$ and $m_{abs} >$ 0.020 $\mu_B$,
20 of them are with $m \approx m_{abs}$ and other 11 with $m_{abs} > m$.


In all 34 structures the distance between the 2  3-fold atoms $>$ 2.19 \AA. 
Since we use a localized basis set with cut-off radius  
of $s$ orbital 2.22 \AA~ (see description in section \ref{sec:energy})
we did one check with longer cut-off radius for one similar sample.
For $r_{cut-off}^{2s}$ = 3.07 \AA~ 
and $r_{cut-off}^{2p}$ = 3.84 \AA~ we get practically the same value of $m$
for the fully relaxed geometry, ruling out the possibility of a numerical artefact. 
The requirement of a minimal distance of 2.19 \AA~ between 3-fold coordinated atoms gives 
613 samples instead of 990, increasing the 
percentage of magnetic samples to $\approx$ 5.5\%.

Within all 613 samples, we  found that the 2  3-fold atoms have either 
0, 1 or 2 common neighbours  but we could not find any correlation between the
number of common neighbours and the presence of magnetic states.
We notice, however, that the presence
of 2 common neighbours means that two 3-fold coordinated atoms form a tetragon. 
Interestingly, such a situation was found 4 times for 
the 34 samples with a magnetic state, a relatively high percentage. 

The above analysis was performed to identify some general geometrical 
configurations leading to magnetism starting from the simplest possible situation. 
Our analysis accounts then for only 34 of the total 202 structures with magnetic moments. 
Nevertheless we have seen that the local environment, namely the 
interatomic distances and coordination of the atoms carrying a magnetic moment, plays a role.
A way to further investigate this point is to compare the 
radial distribution function (rdf) of magnetic atoms with that of all others as 
we do in Fig. \ref{fig:rdf}. 
If we average over all atoms of all 24300 studied configurations, 
the rdf presents two sharp peaks at the interatomic distance 
of graphite (1.42 \AA) and diamond (1.54 \AA) and a broad distribution 
around them from $\sim$ 1.32 \AA~ to $\sim$ 1.7 \AA.
If we consider only the atom with maximum (non-zero) 
magnetic moment within each sample, we find that the two sharp peaks disappear, 
in agreement with our discussion that rules out the possibility of magnetism 
for atoms with only graphitic (fully  $sp^2$) or only diamond(fully  $sp^3$) bonds. 
We see instead a number of peaks in between these two distances 
that have to be related to a mixed bonding. 
We notice that a few peaks are rather pronounced but we did not
manage to assign each of them to a specific bonding configuration. 

In this work we have tried to find out the most
promising configurations giving rise to magnetism. 
As we have discussed the minimal conditions to have magnetic moments are
very stringent and lead to a very small percentage of magnetic structure
among the thousands studied.
Nevertheless some exist and, as we show in the next section 
\ref{sec:exchange-energy}, could lead to magnetic order if suitably repeated.
Moreover, since our analysis only gives a first insight in this phenomenon,
we examine in detail in Section \ref{sec:examples} the structure
of the most interesting realizations found in our quest.

\section{Exchange energy}
\label{sec:exchange-energy}
~

\begin{figure}
    \centering
    \includegraphics[width=0.50\textwidth]{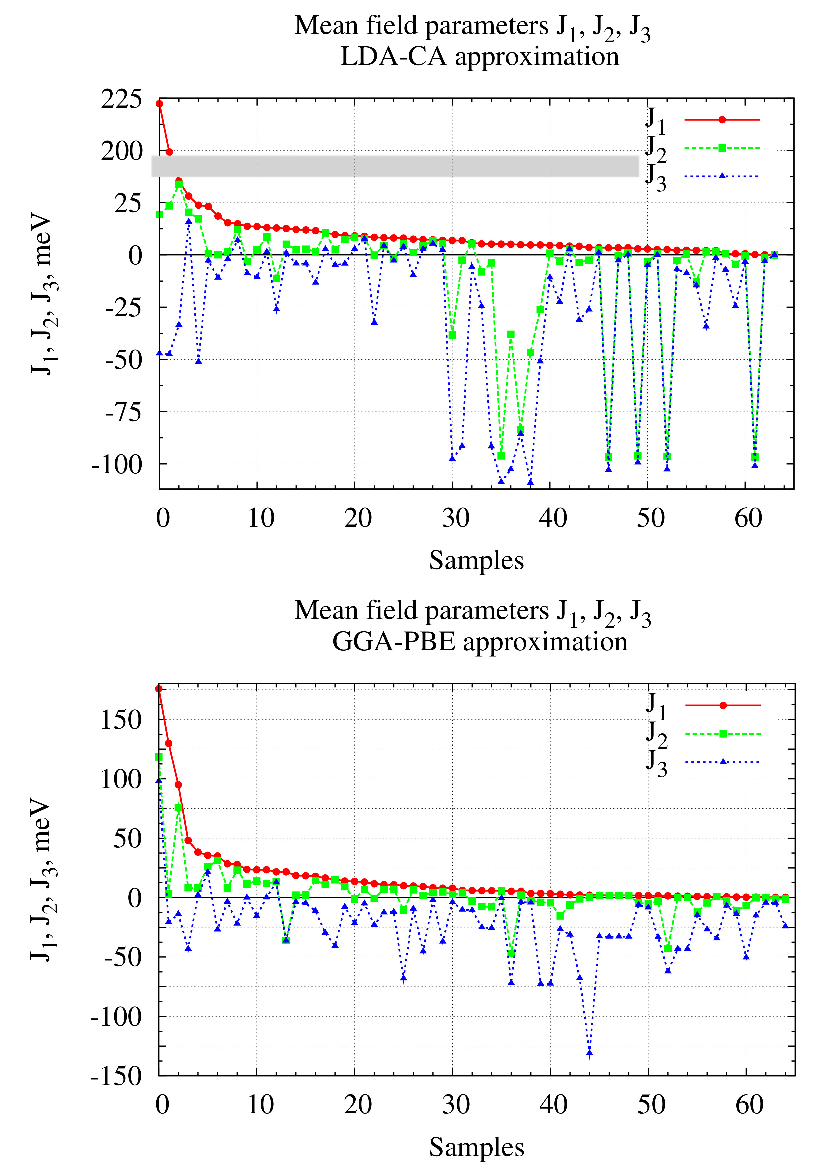}
    \caption{
        Mean field parameters $J_1$, $J_2$, $J_3$ calculated 
        using equation (\ref{eq:J}) in 3 periodical directions 
        sorted by the automatic procedure described in the text. 
        Two approximations were used, i.e. LDA-CA (top panel) and 
        GGA-PBE (bottom panel). Notice the cut in the $Y$ axis of the top panel. 
    }
    \label{fig:J}
\end{figure}

In this section we focus on the samples with atoms carrying a 
magnetic moment and examine the possibility of magnetic order 
in the spirit of mean field theories. 

Most of the studied samples have  orthorhombic unit cells with lattice vectors
$a$, $b$ and $c$ where all three periodic directions 
are different (i.e. anisotropic crystal). 
If the unit cell has only one atom with magnetic moment much larger than 
the others, we calculate the exchange energy $J$ for each
periodic direction separately. 
To do so, we double the original unit cell in the chosen direction and 
initialize the spins on the two magnetic atoms as "up" and "down"
in the first and  second periodic images respectively and calculate the 
total energy which we call here $E_{AFM}$. 
By initializing the spins on the two magnetic atoms as "up" and "up" we calculate the
total energy $E_{FM}$. 
From these two values we calculate the mean field parameters (exchange energy) as 

\begin{equation}
\label{eq:J}
J = \frac {E_{AFM} - E_{FM}} {2}
\end{equation}

Since we have three independent values of $J$ calculated for 
unit cells doubled in the direction of the $a$, $b$ and $c$ lattice vectors, 
we call them here as $J_a$, $J_b$ and $J_c$.
In order to automatize the evaluation and ordering of $J$ even though we 
do not have a priori any knowledge of the studied geometry we use the following procedure.
We start by taking only the samples which have at least one positive value 
among $J_a$, $J_b$, $J_c$. 
Then, for the chosen sample, we sort the values of $J_a$, $J_b$, $J_c$ in descending order.
The sorting breaks the relation with the $a$, $b$ and  $c$ vectors 
so that we rename the sorted values as $J_1$, $J_2$, $J_3$
where $J_1$ is always positive and $J_1 \ge J_2 \ge J_3$.
Finally we group all the chosen samples, sorting them by $J_1$ in descending order 
and make the plot shown in Fig. \ref{fig:J}.
First of all, we notice that the largest values of exchange parameters, both ferromagnetic and antiferromagnetic, 
can reach few hundreds meV. They correspond to magnetic atoms only a few \AA~ apart. 
Such a strong interaction for atoms near to each other is not surprising because, for example, 
a value of hundred meV was found for the ferromagnetic zigzag edges of graphene passivated by Hydrogen\ref{Yazyev2008-edges}.
In our samples that are constructed by periodically repeating relatively small cells, 
this situation occurs a number of times but it is hard to expect it for real carbon foam structures. 

We have to note that, particularly for the large samples, 
and in any case in view of the large number of studied samples,
we could not study the stability of the magnetic 
samples with respect to temperature. We could only perform the
conjugated gradient optimization for the doubled cells.
Both configurations with parallel and antiparallel orientation 
of magnetic moments were checked in this way.

The first 10 points in Fig. \ref{fig:J} with the highest values of
$J$ in LDA-CA approximation, correspond to the following samples: 
AC07-0657, AC07-0680, AC07-0003, AC16-0070, AC07-0907, 
AC14-0388, AC14-0742, AC12-0367, AC15-0776, AC16-1024.
The first 10 points in GGA-PBE approximation correspond to the samples  
AC14-0831, AC07-0907, AC15-0656, AC13-0119, AC15-0267, 
AC10-1116, AC07-0003, AC09-0394, AC07-0657, AC09-0027.
Only the two samples AC07-0657 and AC07-0907 are among the first 10 in both approximations. 
Beside two these samples, the sample AC14-0831 has, in both approximations, a positive value of 
all three $J$'s. 
Unfortunately, as expected, the LDA-CA and GGA-PBE approximations often do not agree,
especially for such a complicated case as disordered carbon where  
changes in the bond length up to 2-3\% can dramatically change 
the  geometry and electronic properties. 
In contrast to planar geometries, like the graphene grain boundaries examined in \cite{DBGB},
these 3D, disordered geometries usually do not have any symmetry 
that can compensate small variations in the bond length.
Despite these shortcomings, two samples with the highest
values of $J$ are present in both approximations.

We can draw some conclusions based on our physical observations.
First of all, the only few samples that have all three values of $J$ positive, 
making it possible to expect 3D ferromagnetism, 
have very small values of $J$ only  up to few meV. 
This finding agrees with the fact that 3D ferromagnetism has been 
experimentally observed in carbon nanofoams only up to 90 K.  

Most samples have two positive values of $J$ (see Fig. \ref{fig:J}) making it possible to have 
2D lattices of magnetic moments with ferromagnetic arrangement. 
In some cases the values of $J$ are rather large and might lead to high Curie temperature. 
Since in 2D ferromagnetic order can exist,
 we pay special attention to this important case.
When the third value of $J$ is negative, the system has antiparallel coupling of
2D ferromagnetic networks. 
To exclude this effect and have only ferromagnetic couplings,
we have to separate the 2D networks  by imposing additional geometrical constrains.
In Fig. \ref{fig:ac-p2} we show how this could be realized by creating 2D grain boundaries.
Such a geometrical isolation of 2D structures could lead to ferromagnetic order in a bulk 3D system.  
In the next section \ref{sec:examples} we discuss the geometry
of the most promising structures, namely those with high values of the $J$ parameters. 

\section{Examples of magnetic structures}~
\label{sec:examples}

%
%

%
%
%
%

%
%

\begin{figure}[ht]
    \centering
    \includegraphics[width=0.20\textwidth]{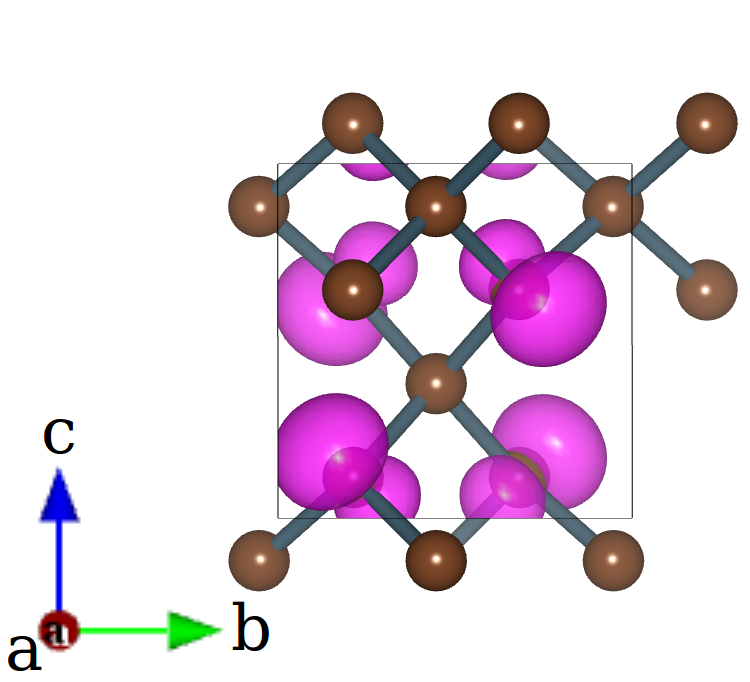}
    \includegraphics[width=0.20\textwidth]{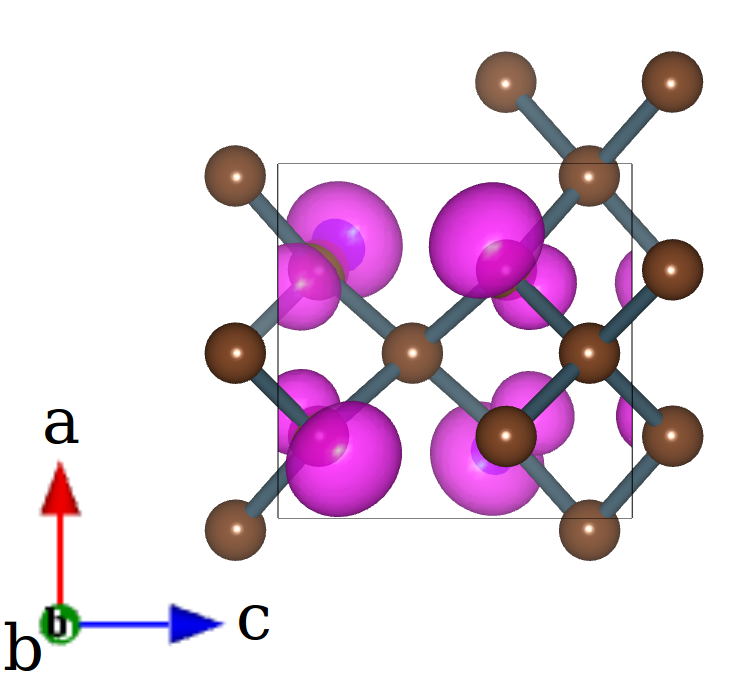}
    \includegraphics[width=0.20\textwidth]{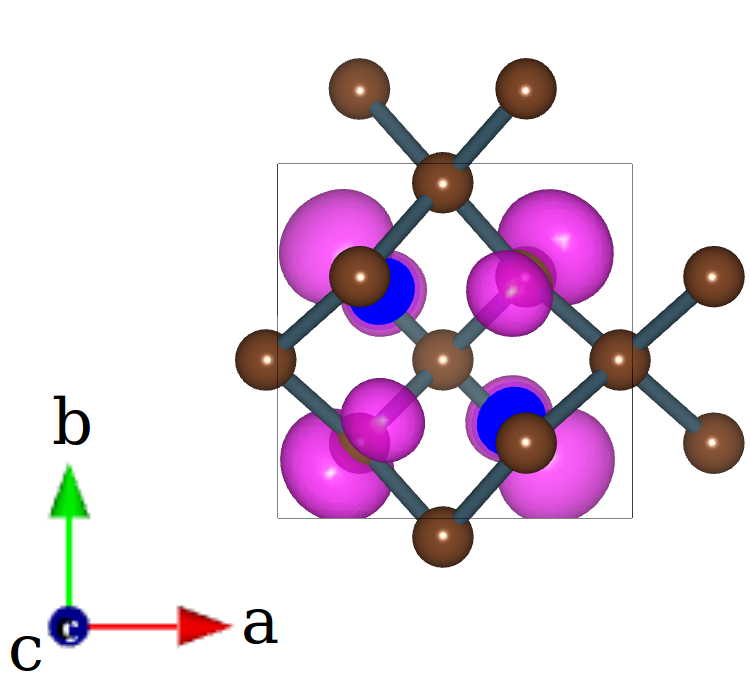}
    \includegraphics[width=0.20\textwidth]{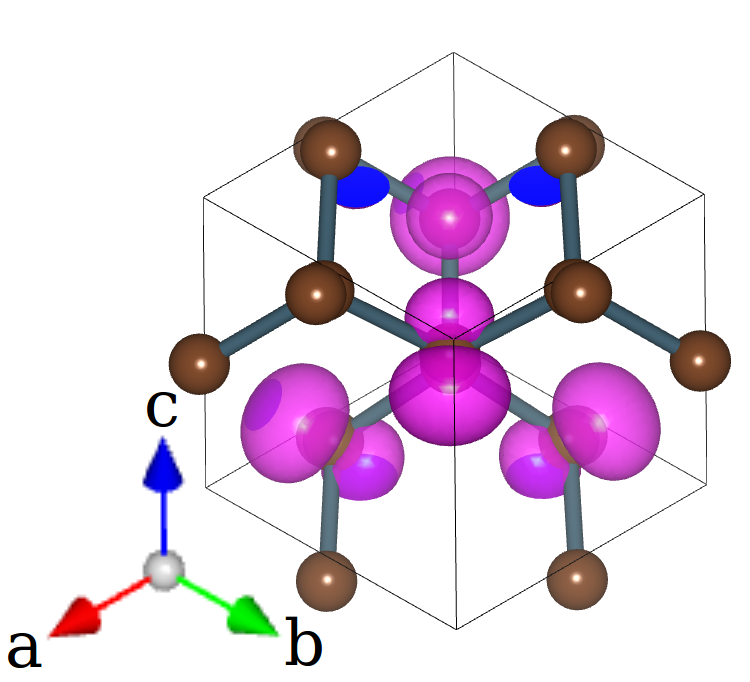}
    \caption{
        Spin density isosurface plot and unit cell of the sample AC07-0010 
        shown (from left to right, top to bottom) in the plane orthogonal 
        to the 100, 010, 001 and 111 directions.
        This structure occurs 11 times in a set of 1000 samples.}
    \label{fig:AC07-0010}
\end{figure}

Here we show a few geometrical examples discovered in our search 
which could be important from different points of view.
Out of many and many different geometries within the series with small number of atoms, 
we will pay attention to the samples AC07-0003 and AC07-0010
(marked as A and B in Fig. \ref{fig:EtotStot}).
Sample AC07-0003  has high values of two $J$ parameters 
(see top-10 list in the text related to Fig. \ref{fig:J}) and was found several
times with small variations in geometry and sample AC07-0010
has a highly symmetric geometry with 4 $sp^2$ atoms
with magnetic moments and  appeared 11 times.

Also we will pay attention to two samples, AC09-0405 and AC14-0831 
(marked as C and D in Fig. \ref{fig:EtotStot}),
due to their low formation energy (left filled region in Fig. \ref{fig:EtotStot},
relatively simple structure, similarity to the geometry of graphite
and ferromagnetic arrangement of magnetic moments.

Finally we briefly discuss two more complicated structures, 
AC09-0708 and AC15-0267, which have 3D ferromagnetic properties 
according to the GGA-PBE approximation.

\begin{figure}[htb]
    \centering
    \includegraphics[width=0.50\textwidth]{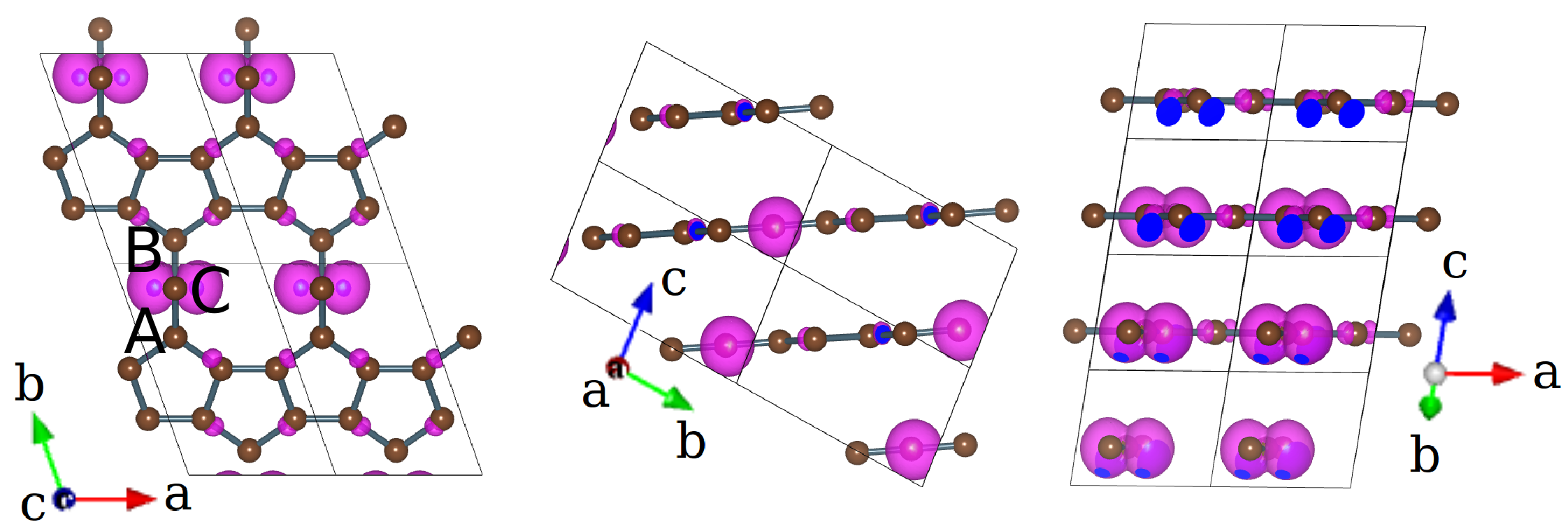}
    \caption{
        Spin density isosurface plot of the sample AC07-0003.
        Top, side and front views.}
    \label{fig:AC07-0003}
\end{figure}

An example of a 3D bipartite unit cell is presented
by the sample AC07-0010 shown in Fig. \ref{fig:AC07-0010}.
Its structure can be described by considering a sequence of
steps starting from a face-centred cubic cell.
Taking into account all 6 atoms sitting at the centre of each face of the unit cell, 
we can construct an octahedron with 6 vertices and 8 faces.
Within the 8 faces, we can choose 4 ones in such a way that any two of them have no common edge. 
Putting 4 atoms at the centre of the 4 chosen faces of the octahedron 
together with the original 3 atoms sitting at the faces of the cubic unit cell
would finally produce our geometry.
The final structure has 7 atoms where 3 of them are 4-fold coordinated 
in $sp^3$ hybridization and 4 atoms are 3-fold coordinated in $sp^2$ hybridization.
Each of the 4 $sp^2$ hybridized atoms carries a magnetic moment shown as a pink cloud
in Fig. \ref{fig:AC07-0010} and contributes the value $m_C=0.061 ~\mu_B$ to the 
total spin polarization. The geometry has 3+4=7 atoms and (3x4+4x3)/2=12 bonds 
each of them of 1.51 \AA, corresponding to a peak in 
the rdf of magnetic atoms in Fig. \ref{fig:rdf}.
This structure presents a frequently occurring pattern where a 3-fold coordinated
carbon atom in the $sp^2$ hybridization is surrounded by 3 $sp^3$ atoms.
In Fig. \ref{fig:bp-graph} we have seen that graphene has a 2D bipartite unit cell
with one atom in the Left and  one in the Right subgraphs.  
The sample AC07-0010 has a 3D bipartite unit cell with 4  $sp^2$ atoms in the Left 
and 3 $sp^3$ atoms in the Right subgraph.
This situation is similar to the structure of half hydrogenated graphene
\cite{hh-graphene} where only the carbon atoms on
one sublattice are $sp^3$ bonded to a hydrogen. 
Exactly this feature allows us to apply the Lieb theorem 
\cite{Lieb-theorem, Yazyev2010-report, KatsnelsonBook}
and expect non zero total spin polarization 
in agreement with the result of our DFT calculations.

\begin{figure*}[htb]
    \centering
    \includegraphics[width=1.00\textwidth]{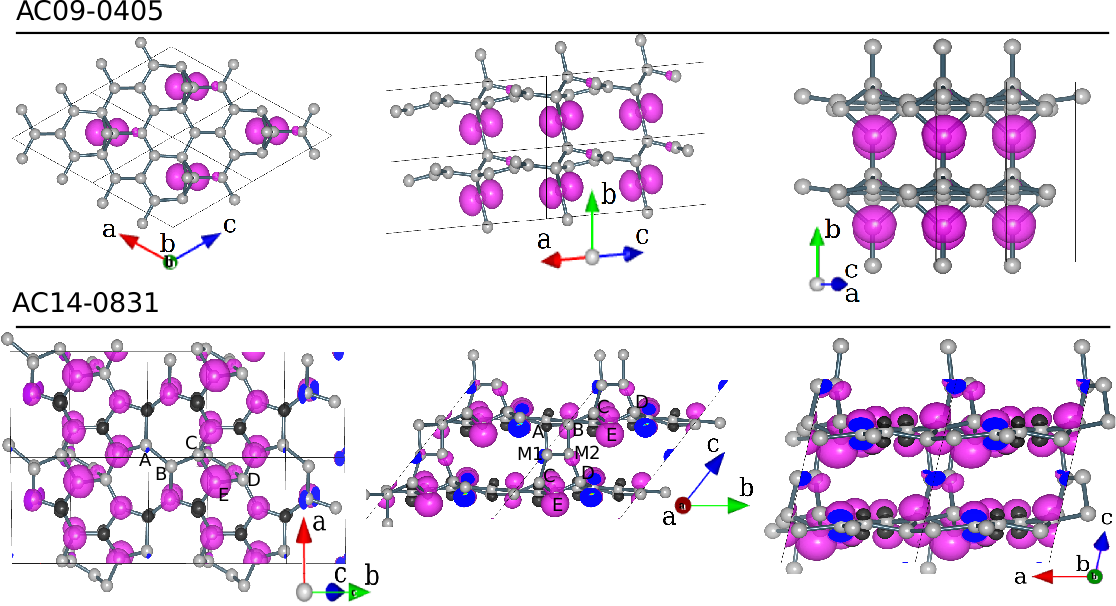}
    \caption{
        Spin density isosurface plot of AC09-0405 and AC14-0831 samples.
        For clarity the unit cell of AC14-0831 was rotated in such a way
        as to put two of the lattice vectors along the graphene plane.}
    \label{fig:ac-p2}
\end{figure*}

The magnetic states in the sample AC07-0003 shown in Fig. \ref{fig:AC07-0003}
are related to the presence of a dangling bond on the 2-fold coordinated 
carbon atom labelled as C in the left panel of Fig. \ref{fig:AC07-0003}.
Beside the rather high values of two mean field parameters $J$'s
(see Fig. \ref{fig:J} and text), such a structure has
another interesting geometrical property.
In Ref.~\cite{Koskinen2008}, the energy of several types of edges was
considered and the reconstructed 5-7 zigzag edges with pentagons and
heptagons were found to have the lowest energy.
Also other types of reconstructed edges were studied and the
armchair 5-6 (pentagon-hexagon) structure was found
to have not much higher energy than the lowest.
The reconstructed 5-6 armchair edge is a structure which can be used 
to construct the geometry of the sample AC07-0003 around the line of 2-fold atoms. To do so we have to take 
two armchair 5-6 edges and connect them together using two 2-fold coordinated 
atoms from the pentagons on each edge 
(atoms A and B in Fig. \ref{fig:AC07-0003} left) with
the addition of an intermediate carbon atom labelled as C.
Such a junction saturates all dangling bonds of atoms A and B while
keeping an unpaired electron on atom C.
The sample AC07-0003 is made of a stacking of such 2D planes with magnetic atoms.
One could describe this structure as a graphite lattice where
the planes contain grain boundaries with 2-fold atoms carrying magnetic moments. 

Another family of structures which can be seen as
modifications of the  graphite structure is
represented  by the two samples AC09-0405 and AC14-0831 both shown
in Fig. \ref{fig:ac-p2}.
The first one, AC09-0405, has a graphite-like structure made of two graphene planes 
separated by an interplanar distance $d_{ll}$ = 3.17 \AA~ 
(to be compared to 3.35 \AA~ in graphite).
The key feature of this geometry is the presence of a 3-fold 
coordinated interstitial atom between the graphene planes with
magnetic moment $m_C = 0.155 ~\mu_B$.
Although this sample has been found in our random search, one can identify a
set of simple geometrical steps to construct it. 
Let us  consider 4 in-plane unit cells of an AA-stacked graphite,
namely a unit cell with 8 atoms and vertical
periodicity equal to the interplane distance.
In the top right panel of Fig. \ref{fig:ac-p2}, 4 such unit cells are shown.
Now select an arbitrary carbon atom in the unit cell and apply the following procedure. 
We replace this atom with two atoms, one above and one below the original plane.
In getting out of plane, these two atoms become close enough to form a bond between planes.
This geometry has to be further relaxed to obtain the final structure
shown in the right panels of Fig. \ref{fig:ac-p2}. 
Once we calculate the spin polarized electronic density, we find that
a large magnetic moment appears on one of the two interstitial atoms.
As shown in the right bottom panel of Fig. \ref{fig:ac-p2}
the magnetic atom is part of a tetragon, a feature that we have discussed
in Section \ref{sec:structure} and that appears in
4 of the 34 magnetic samples with two $sp^2$ atoms.
Possibly, the almost square form of the tetragon,
with angles close to 90 degrees, is of importance.

\begin{figure*}[htb]
    \centering
    \includegraphics[width=1.00\textwidth]{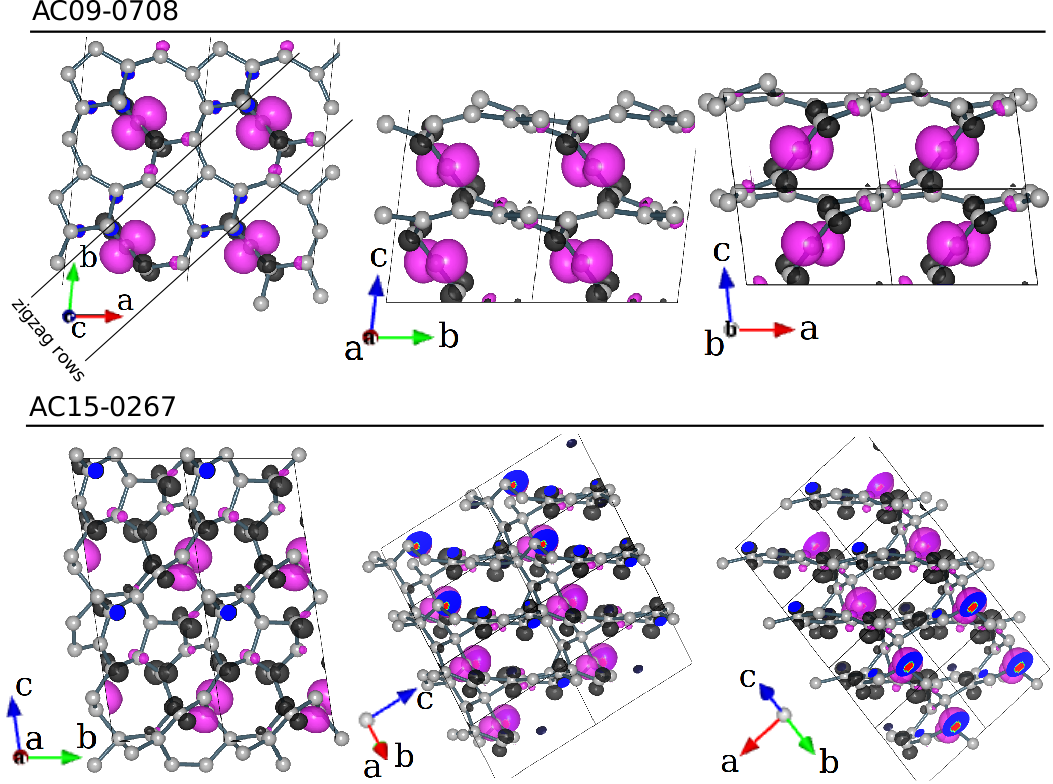}
    \caption{
        Spin density isosurface plot of 
        the samples with 3D ferrimagnetic properties studied 
        within GGA-PBE approximation.}
    \label{fig:ac-p3}
\end{figure*}

The second example of graphite-like structure is AC14-0831 shown in Fig. \ref{fig:ac-p2}.
We can describe its structure as a graphite made of planes of graphene connected
through an interplanar dimer C2 (atoms M1 and M2 in Fig. \ref{fig:ac-p2} left-middle).
Moreover, each graphene plane  has a grain boundary 
\cite{GB-Miller, GB-Coraux, Loginova2009} made of 
a continuous line of Stone-Wales 5-7-7-5 defects \cite{Stone-Wales}.
Another way to imagine the structure of this grain boundary is to join together
two grains with zigzag edges reconstructed as zigzag-57 \cite{Koskinen2008} 
(Fig. \ref{fig:ac-p2} top-left).
The interplanar dimer C2 forms one pentagon and one tetragon with 
bottom and top planes respectively (Fig. \ref{fig:ac-p2} left-middle).
The tetragon is formed by bonds between the C2 dimer and two heptagons 
in the grain boundary in the graphene plane (atoms A and B in Fig. \ref{fig:ac-p2} left-top) 
 while the pentagon is formed by bonds between the C2 dimer and 3 atoms belonging to 
the hexagons in the graphene plane (atoms C, D, E in Fig. \ref{fig:ac-p2} left-top). 
The parallel alignment of the A-B bond to the vector connecting
the atoms C and D gives the possibility of connecting the planes through the
interplanar dimer M1-M2 by forming a pentagon and a tetragon 
with planar arrangement.

The origin of the magnetic states in this configuration is again explained by 
the Lieb theorem \cite{Lieb-theorem, Yazyev2010-report, KatsnelsonBook}
since the  $sp^3$ atoms C and D, belonging to the same sublattice,
break the bipartite lattice symmetry.
This situation is similar to the one encountered in
the junction of a nanotube with a graphene ribbon\cite{ribbon-tube}.
Also there, the 4-fold atoms brake the bipartite symmetry,
leading to magnetic moments. 

The last set of samples discovered in our search has a
3D ferro/ferri-magnetic structure with large values of $J$.
The samples AC16-0070 and AC15-0776 studied by LDA-CA
(point N4 and N9 in Fig. \ref{fig:J} top) have a complicated geometry that
we do not show here because it cannot be described in
a simplified way as done previously.
Within the samples studied by GGA-PBE
we select AC15-0267, AC10-1116 and AC09-0708 
(4th, 5th and 12-th points  in Fig. \ref{fig:J} bottom) and 
show two of them in Fig. \ref{fig:ac-p3}.
The sample AC15-0267 can be classified as a graphite-like geometry
with modifications in the graphene plane
(i.e. grain boundaries and/or point defects) and interplane atoms
(see Fig. \ref{fig:ac-p3} left-middle) similar to the geometry of AC14-0831. 
One more example of magnetic states due to the presence of dangling bonds
is given by the sample AC09-0708 shown in the right panels of Fig. \ref{fig:ac-p3}. 
This sample also has a layered structure where the 2-fold coordinated atoms 
always  connect two layers. Each layer itself is made of
two pentagons and two octagons arranged along a line between two zigzag rows.
A top view of the sample gives an image very similar to
the high angle grain boundary found for graphene on Ni in ref. \cite{Lahiri}.

All structures presented in this section are available\cite{ac-xsf-data}
in the xsf file format suitable for use in the VESTA program \cite{vesta}.

\section{Conclusions}~
\label{sec:conclusion}

In this work we have presented the results of a massive, 
automated search of disordered carbon structures with magnetic states. 
We have tried to identify the common structural features present in 
the samples where magnetic moments appear. 
In our analysis we have used elements of graph theory and 
the analogy to structural motifs in graphene grain boundaries. 
We believe that our computational approach can lead to 
progress in the understanding of $s-p$ electron magnetism.  
This task is however still far from complete and the ways to create 
magnetic order are still elusive and will require further investigations.

~
\section*{Acknowledgments}
The support by the Stichting Fundamenteel Onderzoek der Materie (FOM) and
the Netherlands National Computing Facilities foundation (NCF) are
acknowledged. 
~

\bibliographystyle{iopart-num}
\bibliography{bibliography}


\end{document}